\documentclass[aps,twocolumn, noshowpacs, floatfix]{revtex4}
\usepackage{}

\usepackage{amsfonts}
\usepackage{amssymb}
\usepackage{graphicx}
\usepackage{amsmath}
\usepackage[english]{babel}
\usepackage{color}

\begin{document}

\title{Transient dynamics of a one-dimensional Holstein polaron under the influence of an external electric field}

\author{Zhongkai Huang, Lipeng Chen, Nengji Zhou, and Yang Zhao\footnote{Electronic address:~\url{YZhao@ntu.edu.sg}}}
\affiliation{Division of Materials Science, Nanyang Technological University, Singapore 639798, Singapore}

\begin{abstract}
Following the Dirac-Frenkel time-dependent variational principle, transient dynamics of a one-dimensional Holstein polaron with diagonal and off-diagonal exciton-phonon coupling in an external electric field is studied by employing the multi-D$_2$ {\it Ansatz}, also known as a superposition of the usual Davydov D$_2$ trial states.  Resultant polaron dynamics has significantly enhanced accuracy, and is in perfect agreement with that derived from the hierarchy equations of motion method. Starting from an initial broad wave packet, the exciton undergoes typical Bloch oscillations. Adding weak exciton-phonon coupling leads to a broadened exciton wave packet and a reduced current amplitude. Using a narrow wave packet as the initial state, the bare exciton oscillates in a symmetric breathing mode, but the symmetry is easily broken by weak coupling to phonons, resulting in a non-zero exciton current. For both scenarios, temporal periodicity is unchanged by exciton-phonon coupling. In particular, at variance with the case of an infinite linear chain, no steady state is found in a finite-sized ring within the anti-adiabatic regime. For strong diagonal coupling, the multi-$\rm D_2$ {\it Anstaz} is found to be highly accurate, and the phonon confinement gives rise to exciton localization and decay of the Bloch oscillations.
\end{abstract}
\date{\today}

\maketitle

\section{Introduction}

Early evidence of time-domain Bloch oscillations (BOs) in superlattices was revealed by transient degenerate four-wave mixing (DFWM) \cite{fe_92, me_93},
yielding results in agreement with the work of Esaki and Tsu \cite{es_70}. BOs emerge when an electron subject to a perfectly periodic lattice potential executes periodic motion in the presence of an external electric field \cite{bl_28, ze_34, ch_83, wa_59, ho_40, Bloch1, Bloch2}. In recent decades, BOs were found in various physical systems. For example, with the progress of laser cooling and manipulation, cold atoms in an optical lattice were found to exhibit BOs and Wannier-Stark ladders \cite{da_87, ch_91}, in accordance with theoretical predictions \cite{da_96, wi_96, mo_01}.
Further experimental evidence of BOs was found in atomic Bose-Einstein condensates (BEC) in optical lattices \cite{an_98, ko_10, cr_02}, in a semiconductor waveguide \cite{mo_99}, and in a thermo-optic polymer array subjected to a temperature gradient \cite{pe_99}. It was also suggested that waveguide arrays with a changeable effective index of the individual guides would be an optimal system to detect optical BOs in the space domain \cite{pe_98}. In addition, BOs have received much attention in the past decades for its potential application in THz generation and negative differential conductance \cite{es_70}. Investigations of BOs have been carried out in THz emission \cite{wa_93}, electro-optic detection \cite{fo_2000}, and DFWM experiments \cite{fe_92}. Recently, a new mesoscopic amplifier by the name of Bloch Oscillating Transistor has been proposed based on BOs \cite{de_03}.

Following the remarkable experimental progress, recent theoretical work explores the presence of BOs in a variety of contexts. For example, formation of photonic BOs was investigated in an exponentially chirped one-dimensional Bragg grating using Hamiltonian optics, where paths of geometrical rays are determined from Hamilton's equations \cite{wi_02}. BOs were also theoretically predicted to exist in magnetic systems, such as soliton-like domain walls in anisotropic spin $1/2$ chains under magnetic fields \cite{kr_98}. Furthermore, BOs in interacting quantum few body systems have been modeled with the Bose-Hubbard model \cite{bu_03}.

One important issue for BOs dynamics is the effect of carrier-phonon interactions, which is of essential importance for systems such as semiconductor superlattices \cite{dekorsy_2000, ghosh_2000}, organic materials \cite{jo_01, ba_02, jo_04, qu_09, pe_89, di_08, Rakhmanova_1999} and quantum dots \cite{dmitriev_2001, petroff_2001}. The phonon bath is treated classically and the associated carrier-phonon coupling semi-classically in the Su-Schrieffer-Heeger model for polymers \cite{jo_01, ba_02, jo_04, qu_09} and in the Peyrard-Bishop-Holstein model for DNA \cite{pe_89, di_08}. More than forty years ago, Thornber and Feynman studied the motion of an electron in a polar crystal in a strong electric field using Fr\"ohlich's model of polaron, and found that the electron acquires a constant velocity due to the emission of phonons \cite{th_70}. Much theoretical work based on the rate equations or the Boltzmann equation later focused on calculating transition rate probabilities, rather than complex quantum amplitudes \cite{kummel_1978, Emin1987}. However, full quantum coherence was revealed to be retained in an inelastic quantum transport process and a steady state was found to be reached subject to an infinite lattice, leaving the dispersionless optical phonon absorbing the excess energy from the external field \cite{bo_97, vi_11}.

Despite dedicated studies of BOs dynamics, the effect of complex interplay between the electron and its accompanying phonon cloud in a one-dimensional lattice remains an open question \cite{zhang_02, cheung_13}.
Spatial dynamics influenced by exciton-phonon coupling is also inadequately studied given initial broad and narrow exciton wave packets \cite{ly_1997, su_98}. Moreover, a unified treatment on various types of exciton-phonon coupling remains elusive.

The intramolecular (diagonal) and intermolecular (off-diagonal) exciton-phonon coupling has been demonstrated to coexist in organic materials \cite{coropceanu_07}, and it is shown that off-diagonal coupling plays a crucial role in polaronic diffusion \cite{tamura_2012}. In the presence of an external electric field, polaron motion with off-diagonal coupling in polymer chains has been simulated with a semiclassical method \cite{Rakhmanova_1999, Johansson_2002}, neglecting the quantum nature of phonons and exciton-phonon coupling. Fully quantum mechanical treatments are few in the literature with little attention paid to Hamiltonians with off-diagonal exciton-phonon coupling and an external field  due to inherent difficulties to obtain reliable solutions \cite{vi_11}.

Based on the Holstein molecular crystal model, which describes the motion of an exciton enveloped by a cloud of phonons \cite{ho_59}, we include both diagonal and off-diagonal exciton-phonon coupling. Despite apparent simplicity of its Hamiltonian, the Holstein model never ceases to surprise us with rich physics related to exciton-phonon correlations. The off-diagonal coupling was investigated earlier by the Munn-Silbey theory \cite{mu_85} and a variational method using the Davydov D$_2$ {\it Ansatz} \cite{zh_12} and the global-local {\it Ansatz} \cite{zh_97}. Recently, Zhao {\it et al.}~developed a refined trial state, the multiple Davydov D$_2$ {\it Ansatz,} to accurately treat the dynamics of the Holstein model with simultaneous diagonal and off-diagonal coupling \cite{zh_15, zhou_2016}. In this work we will apply this method to polaron dynamics in an external field.

In addition, only interactions between the exciton and optical phonons were considered previously \cite{th_70, vi_11, zh_15}. Quantum-mechanical calculations indicated that excitons are coupled to both acoustic and optical phonons \cite{bredas_2012, li_2013}. Discussions on the role played by acoustic modes are inadequate in the literature, especially in the simultaneous presence of acoustic and optical phonons accompanied by an external field \cite{davydov_1967, la_1997}.

In this work, we investigate the impact of diagonal and off-diagonal exciton-phonon coupling on polaron dynamics in a ring system in the presence of a constant electric field using the multiple Davydov D$_2$ trial state with the Dirac-Frenkel variational principle. Accuracy of the method is verified by the comparison with benmark calculations obtained from the numerically exact hierarchical equations of motion (HEOM) method \cite{Tanimura1, ch_15} which is also firstly used in this work to deal with field-driven cases to the best of our knowledge. The validity of our variational method is also carefully examined by quantifying how faithfully our result follows the Schr\"odinger equation in balance with the computational efficiency. For weak exciton-phonon coupling in the anti-adiabatic regime, we consider both acoustic and optical phonons, and calculate the time evolution of various quantities with special attention paid to the effect of weak dissipation on the spatial amplitudes of BOs using different initial conditions. Treating strong diagonal coupling is a formidable challenge for the HEOM method, but our variational method using the multi-$\rm D_2$ {\it Ansatz} remains numerically affordable (in fact, with a higher precision with stronger coupling).

The rest of the paper is structured as follows. In Sec.~\ref{methodology}, we present the Hamiltonian in the gauge transformed form and our trial
wave function, the multi-$\rm D_2$ {\it Ansatz}. In Sec.~\ref{Validity of variational dynamics}, our variational results are compared to those of the benchmark HEOM calculations.
In Sec.~\ref{Weak and intermediate diagonal coupling}, the
influence of weak diagonal coupling in the anti-adiabatic regime on the exciton wave packet is investigated using initial Gaussian wave packets with varying widths with particular attention paid to the possible existence of a steady state. Finally, polaron dynamics is examined in  the strong diagonal coupling regime in
Sec.~\ref{Strong diagonal coupling}. Conclusions are drawn in Sec.~\ref{Conclusions}.

\section{methodology}
\label{methodology}
\subsection{Model}
\label{Model}
The Hamiltonian of the one-dimensional Holstein polaron reads
\begin{equation}
\hat{H}=\hat{H}_{\rm ex}+\hat{H}_{\rm ph}+\hat{H}_{\rm ex-ph}^{\rm diag}+\hat{H}_{\rm ex-ph}^{\rm o.d.}
\label{Holstein}
\end{equation}
where $\hat{H}_{\rm ex},\hat{H}_{\rm ph}$,$\hat{H}_{\rm ex-ph}^{\rm diag}$ and $\hat{H}_{\rm ex-ph}^{\rm o.d.}$ denote the exciton Hamiltonian, the bath
(phonon) Hamiltonian, the diagonal and off-diagonal exciton-phonon coupling Hamiltonian, respectively. An extra term {\it $F\sum_{n}na_{n}^{\dagger}a_{n}$} is
added to represent the scalar potential induced by a constant external electric field $F$. After bending a linear chain of atoms into a ring, however,
potential of the field would become discontinuous at end points, leaving the use of periodic boundary conditions questionable \cite {za_68}. A later
treatment \cite{kr_86} using a gauge transformed vector potential \cite{ki_63} avoids this problem and is thus suitable for ring systems. We will use the gauge transformed Hamiltonian for the ring system under a constant electric field (see Supporting Information for details), which can be defined as
\begin{eqnarray}\label{Hamiltonian_site}
\hat{H}_{\rm ex} & = & -J\sum_{n}a_{n}^{\dagger}\left(e^{-iFt}a_{n+1}+e^{iFt}a_{n-1}\right), \nonumber \\
\hat{H}_{\rm ph}&=&\omega_{0}\sum_{n}b_{n}^{\dagger}b_{n},\nonumber \\
\hat{H}_{\rm ex-ph}^{\rm diag}&=&-g\omega_{0}\sum_{n}a_{n}^{\dagger}a_{n}\left(b_{n}+b_{n}^{\dagger}\right), \nonumber\\
\hat{H}_{\rm ex-ph}^{\rm o.d}	&=&	\frac{1}{2}\phi\omega_{0}\sum_{n,l}\left[a_{n}^{\dagger}a_{n+1}\left(b_{l}+b_{l}^{\dagger}\right)\left(\delta_{n+1,l}-\delta_{n,l}\right)\right.\nonumber \\
&  &+\left.a_{n}^{\dagger}a_{n-1}\left(b_{l}+b_{l}^{\dagger}\right)\left(\delta_{n,l}-\delta_{n-1,l}\right)\right]
\end{eqnarray}
where $\hat{a}_n^{\dagger}$ ($\hat{a}_n$) and $\hat{b}_n^{\dagger}$ ($\hat{b}_n$) are the exciton and phonon creation (annihilation) operators for the $n$-th site, respectively. Written in the phonon momentum space,
\begin{eqnarray}\label{Hamiltonian}
\hat{H}_{\rm ph}&=&\sum_{q}\omega_{q}\hat{b}_{q}^{\dagger}\hat{b}_{q},   \\
\hat{H}_{\rm ex-ph}^{\rm diag}&=& -N^{-1/2}g\sum_{n,q}\omega_{q}\hat{a}_{n}^{\dagger}\hat{a}_{n}\left(e^{iqn}\hat{b}_{q}+e^{-iqn}\hat{b}_{q}^{\dagger}\right),\nonumber\\
\hat{H}_{\rm ex-ph}^{\rm o.d.}&=& \frac{1}{2}N^{-1/2}\phi\sum_{n,q}\omega_{q}\{\hat{a}_{n}^{\dagger}\hat{a}_{n+1}[e^{iqn}(e^{iq}-1)\hat{b}_{q}+ {\rm H.c.}]
\nonumber \\
 &  & +\hat{a}_{n}^{\dagger}\hat{a}_{n-1}[e^{iqn}(1-e^{-iq})\hat{b}_{q}+ {\rm H.c.}]\}, \nonumber
\end{eqnarray}
where $\omega_q$ is the phonon frequency with momentum $q$, and $\hat{b}_q^{\dagger}$ ($\hat{b}_q$) is the creation (annihilation) operator of a phonon with momentum $q$,
\begin{equation}
\hat{b}_q^{\dagger} = N^{-1/2}\sum_n e^{iqn}\hat{b}_n^{\dagger}, \quad \hat{b}_n^{\dagger} =  N^{-1/2}\sum_q e^{-iqn}\hat{b}_q^{\dagger}.
\label{momentum}
\end{equation}
The parameters $J$, $g$ and $\phi$ represent the transfer integral, the diagonal coupling strength and off-diagonal coupling strength, respectively.
In this work, we take $\omega_{q}=\omega_{0}\left|\sin\left(q/2\right)\right|$ as the dispersion relation for acoustic phonons, where $q=2\pi l/N$ represents the momentum index with $l=-\frac{N}{2}+1, \ldots, \frac{N}{2}$; in the case of optical phonons, we consider the Einstein dispersionless model, i.e., $\omega_{q}=\omega_{0}$. For simplicity, the Debye frequency is taken to be equal to the Einstein frequency $\omega_{0}$.
In the remainder of the paper, $\omega_0$ is set to unity as the energy unit.

\subsection{Multiple Davydov trial states}

\label{Multiple Davydov trial state}

The multiple Davydov trial states with multiplicity $M$, which are essentially $M$ copies of the corresponding single Davydov {\it Ansatz}, have been developed to investigate the time evolution of the Holstein polaron following the Dirac-Frenkel variational principle. The multi-${\rm D}_2$ {\it Ansatz} has less variational variables than the multi-${\rm D}_1$ {\it Ansatz} when same $M$ is used, nonetheless it can more accurately describe polaron dynamics in the presence of off-diagonal coupling \cite{zh_15}. In this work, we employ the multi-${\rm D}_2$ {\it Ansatz}. It can be constructed as
\begin{eqnarray}\label{D2_state}
&& \left|{\rm D_2^M}\left(t\right)\right\rangle  =  \sum_{i}^{M}\sum_{n}^N\psi_{in}\left|n\right\rangle \left|\lambda_{i}\right\rangle, \\ \nonumber
&& =\sum_{i}^{M}\sum_{n}^N\psi_{in} \hat{a}_{n}^{\dagger}\left|0\right\rangle _{\rm ex} \exp\left\{
\sum_{q}\left[\lambda_{iq}\hat{b}_{q}^{\dagger}-\lambda_{iq}^{\ast}\hat{b}_{q}\right]\right\} \left|0\right\rangle _{\rm ph},
\end{eqnarray}
where $\psi_{in}$ and $\lambda_{iq}$ are time-dependent variational parameters for the exciton probability and phonon displacement, respectively, $n$ represents the site
number, and $i$ labels the coherent superposition state. If $M=1$, the $|\rm D_2^M(t)\rangle$ {\it Ansatz}
is restored to the usual Davydov $\rm D_2$ trial state. The equations of motion of the variational
parameters $\psi_{in}$ and $\lambda_{iq}$ are then derived by adopting the Dirac-Frenkel variational principle,
\begin{eqnarray}\label{eq:eom1}
\frac{d}{dt}\left(\frac{\partial L_2}{\partial\dot{\psi_{in}^{\ast}}}\right)-\frac{\partial L_2}{\partial\psi_{in}^{\ast}} & = & 0, \nonumber \\
\frac{d}{dt}\left(\frac{\partial L_2}{\partial\dot{\lambda_{iq}^{\ast}}}\right)-\frac{\partial L_2}{\partial\lambda_{iq}^{\ast}} & = & 0.
\end{eqnarray}
for the multi-$\rm D_2$ {\it Ansatz}, and the Lagrangian $L_2$ is formulated as
\begin{eqnarray}
L_2 & = & \langle {\rm D}^M_2(t)|\frac{i}{2}\frac{\overleftrightarrow{\partial}}{\partial t}- \hat{H}|{\rm D}^M_2(t)\rangle \nonumber \\
& = & \frac{i}{2}\left[ \langle {\rm D}^M_2(t)|\frac{\overrightarrow{\partial}}{\partial t}|{\rm D}^M_2(t)\rangle - \langle {\rm
D}^M_2(t)|\frac{\overleftarrow{\partial}}{\partial t}|{\rm D}^M_2(t)\rangle \right] \nonumber \\
&-& \langle {\rm D}^M_2(t)|\hat{H}|{\rm D}^M_2(t)\rangle.
\label{Lagrangian_2}
\end{eqnarray}
where the first term yields
\begin{eqnarray}
 &  & \frac{i}{2}\left[ \langle {\rm D}^M_2(t)|\frac{\overrightarrow{\partial}}{\partial t}|{\rm D}^M_2(t)\rangle - \langle {\rm
 D}^M_2(t)|\frac{\overleftarrow{\partial}}{\partial t}|{\rm D}^M_2(t)\rangle \right] \nonumber \\
 &  &= \frac{i}{2}\sum_{i,j}^{M}\sum_{n}\left(\psi_{jn}^{\ast}\dot{\psi}_{in}-\dot{\psi}_{jn}^{\ast}\psi_{in}\right)S_{ji}\nonumber \\
 &  &
 +\frac{i}{2}\sum_{i,j}^{M}\sum_{n}\psi_{jn}^{\ast}\psi_{in}S_{ji}\sum_{q}\left[\frac{\dot{\lambda}_{jq}^{\ast}\lambda_{jq}+\lambda_{jq}^{\ast}\dot{\lambda}_{jnq}}{2}\right.\nonumber
 \\
 &  &
 \left.-\frac{\dot{\lambda}_{iq}\lambda_{iq}^{\ast}+\lambda_{iq}\dot{\lambda}_{iq}^{\ast}}{2}+\lambda_{jq}^{\ast}\dot{\lambda}_{iq}-\lambda_{iq}\dot{\lambda}_{jq}^{\ast}\right],
\label{energies}
\end{eqnarray}
and the second term takes the form
\begin{eqnarray}
 &  & \left\langle {\rm D}^M_{2}\left(t\right)\right|\hat{H}\left|{\rm D}^M_{2}\left(t\right)\right\rangle \nonumber \\
 &  &= \left\langle {\rm D}^M_{2}\left(t\right)\right|\hat{H}_{\rm ex}\left|{\rm D}^M_{2}\left(t\right)\right\rangle +\left\langle {\rm
 D}^M_{2}\left(t\right)\right|\hat{H}_{\rm ex}\left|{\rm D}^M_{2}\left(t\right)\right\rangle \nonumber \\
 &  & +\left\langle {\rm D}^M_{2}\left(t\right)\right|\hat{H}_{\rm ex-ph}^{\rm diag}\left|{\rm D}^M_{2}\left(t\right)\right\rangle+\left\langle {\rm
 D}^M_{2}\left(t\right)\right|\hat{H}_{\rm ex-ph}^{\rm o.d.}\left|{\rm D}^M_{2}\left(t\right)\right\rangle, \nonumber \\
 \end{eqnarray}
where the Debye-Waller factor is formulated as
\begin{equation}
S_{ij} = \exp\left\{ \sum_{q}\lambda_{iq}^{*}\lambda_{jq}-\frac{1}{2}\left(\left|\lambda_{iq}\right|^{2}+\left|\lambda_{jq}\right|^{2}\right)\right\} .
\label{A2}
\end{equation}

Detailed derivations of the equations of motion for the variational parameters are given in Supporting Information. In the numerical calculations, the
fourth-order Runge-Kutta method is used to integrate the equations of motion.

With the wave function $|{\rm D}^M_2(t)\rangle$ available, the total energy
$E_{\rm total} = E_{\rm ex}+ E_{\rm ph}+ E_{\rm diag}+ E_{\rm off}$ is calculated, where
$E_{\rm ex} = \langle {\rm D}^M_{2}|\hat{H}_{\rm ex}|{\rm D}^M_{2} \rangle, ~ E_{\rm ph} = \langle {\rm D}^M_{2}|\hat{H}_{\rm ph}|{\rm D}^M_{2} \rangle,
 ~ E_{\rm diag} = \langle {\rm D}^M_{2}|\hat{H}_{\rm ex-ph}^{\rm diag}|{\rm D}^M_{2} \rangle$, and $E_{\rm off} = \langle {\rm D}^M_{2}|\hat{H}_{\rm
 ex-ph}^{\rm o.d.}|{\rm D}^M_{2} \rangle$.  Additionally, time evolution of the exciton probability $P_{\rm ex}(t, n)$ and the phonon displacement
 $X_{\rm ph}(t, n)$ is also calculated
\begin{eqnarray}
P_{\rm ex}(t, n) &=& \langle {\rm D}^M_{2} |\hat{a}_{n}^\dagger\hat{a}_{n}|{\rm D}^M_{2}\rangle, \nonumber \\
X_{\rm ph}(t, n) &=& \langle {\rm D}^M_{2} |\hat{b}_{n}+\hat{b}_{n}^\dagger|{\rm D}^M_{2}\rangle.
\label{exciton-phonon}
\end{eqnarray}
We further use a standard deviation $\sigma(t)$ and a mean value $c(t)$ of the exciton wave packet to characterize the motion of the exciton wave, which are
defined as follows
\begin{eqnarray}
c(t)& = &\sum_{i}^{N}nP_{\rm ex}(t, n) \nonumber \\
{\sigma(t)}^2 & = &	\sum_{i}^{N}\left(n-c(t) \right)^{2}P_{\rm ex}(t, n)
\label{exciton-phonon}
\end{eqnarray}
where the mean value $c(t)$ describes the centroid of the exciton wave packet, and the standard deviation is used to measure how far the exciton wave spreads
out from the mean position. It is noted that the standard deviation $\sigma(t)$ and the mean value $c(t)$ are sensitive to the initial standard deviation $\sigma_0$ of the exciton wave packet.

\section{Numerical results and discussions}
\label{Numerical results and discussions}
\subsection{Validity of variation for transient dynamics}
\label{Validity of variational dynamics}
In this subsection, we show that the multi-$\rm D_2$ {\it Anstaz} with sufficiently large multiplicity $M$ yields quantitatively accurate solutions to
the dynamics of the Holstein polaron with both diagonal and off-diagonal coupling, in perfect agreement with the benchmark calculations of the
numerically exact HOEM method (see Sec. $4$ of Supporting Information) \cite{Tanimura1, ch_15}. For simplicity, only optical phonons are used in this subsection.

\subsubsection{Diagonal Coupling}
\label{Diagonal Coupling}

\begin{figure}[tbp]
\centering
\includegraphics[scale=0.4]{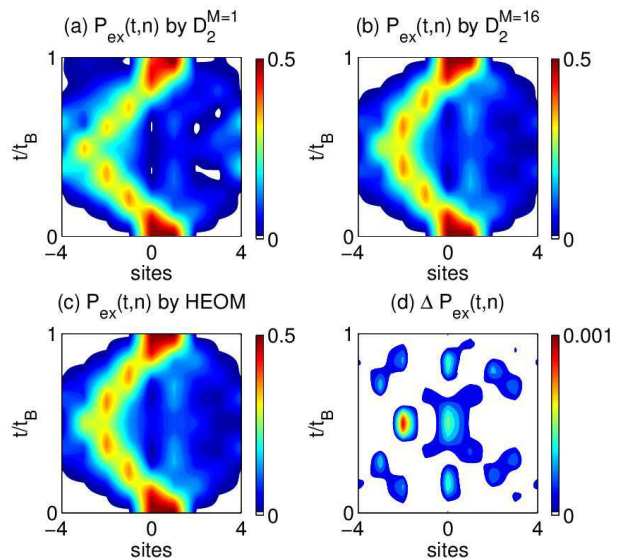}
\caption{Time evolution of the exciton probability $P_{\rm ex}(t, n)$ for a diagonal coupling case of $J=0.1, g=0.28$ and $F=0.1$ is obtained from (a) the single
$\rm D_2^{M=1}$ {\it Ansatz}, (b) the $\rm D_2^{M=16}$ {\it Ansatz}, and (c) the HEOM method. The difference $\Delta P_{\rm ex}(t, n)$ between the HEOM
and the $\rm D_2^{M=16}$ trial state is displayed in (d). The time unit $t_B$ denotes the time period of BOs. $N=8$ is used in the calculations.}
\label{exm1m16heom}
\end{figure}
First, we study the case of diagonal exciton-phonon coupling. Dynamics of the Holstein polaron under a constant external field is investigated by using the multi-$\rm D_2$ {\it Anstaz}, in comparison
with those obtained with the single Davydov D$_2$ {\it Ansatz} and the HEOM method. Using these approaches, the time evolution of the exciton
probability $P_{\rm ex}(t, n)$ as shown in Fig.~\ref{exm1m16heom} is simulated in the case of $J=0.1, g=0.28$ and $F=0.1$ in a small ring with $N=8$ sites for simplicity. The exciton is created on two nearest neighboring sites $\psi_n=(\delta_{n,N/2}+\delta_{n,N/2+1})/\sqrt2$, and the phonon displacement $\lambda_{i,q}(t=0)=0$ is set. As depicted in Figs.~\ref{exm1m16heom}(a) and (b), distinguishable deviations in $P_{\rm ex}(t, n)$ can be found between the variational results from the $\rm D^{M=1}_2$ and $\rm D^{M=16}_2$ {\it Ans\"atze}. Interestingly, $P_{\rm ex}(t, n)$ obtained from the HEOM method in Fig.~\ref{exm1m16heom}(c) is almost identical to that by the $\rm D_2^{M=16}$ {\it Ansatz} in Fig.~\ref{exm1m16heom}(b). Furthermore, the difference in the time evolution of the exciton probability between the variational and HEOM methods, $\Delta P_{\rm ex}(t, n)$, as displayed in Fig.~\ref{exm1m16heom}(d), is two orders of magnitude smaller than the value of $P_{\rm ex}(t, n)$, indicating that the variational dynamics of the Holstein polaron under the external field can be numerically exact if the multiplicity $M$ of the multi-$\rm D_2$ {\it Ansatz} is sufficiently large.

It should be noted that the HEOM method is numerically expensive and thus impractical when the system size is large, while time dependent variational
approaches are still valid to treat the polaron dynamics for large systems once a proper trial wave function is adopted.

\subsubsection{Off-diagonal Coupling}
\label{Off-diagonal Coupling}
\begin{figure}[tbp]
\centering
\includegraphics[scale=0.4]{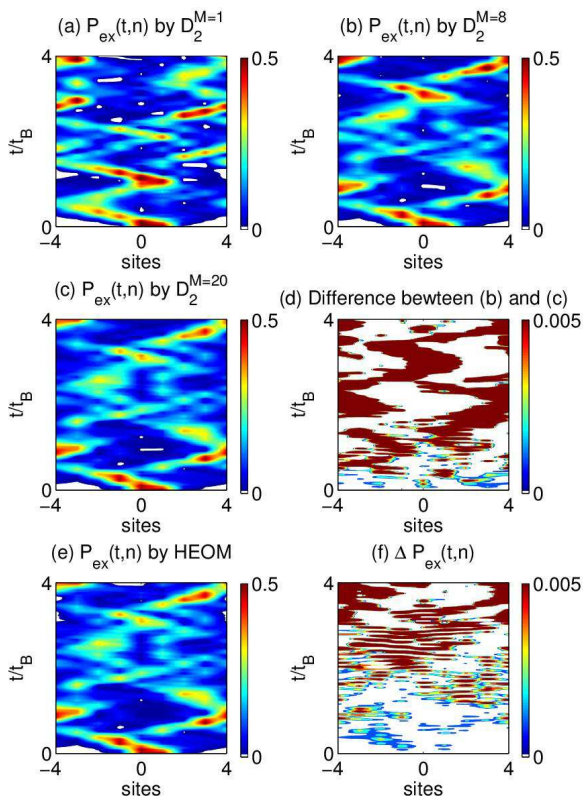}
\caption{Time evolution of the exciton probability $P_{\rm ex}(t, n)$ for a off-diagonal coupling case of $J=0.1, \phi=0.28$ and $F=0.1$ is obtained from (a) the
$\rm D_2^{M=1}$ {\it Ansatz}, (b) the $\rm D_2^{M=8}$ {\it Ansatz}, (c) the $\rm D_2^{M=20}$ {\it Ansatz}, and (e) the HEOM method. (d) The difference
between the $\rm D_2^{M=8}$ and $\rm D_2^{M=20}$ {\it Ans\"atze} and (f) $\Delta P_{\rm ex}(t, n)$ between the HEOM and the $\rm D_2^{M=20}$  {\it Ansatz} are displayed.}
\label{exN8m1m8m20heom}
\end{figure}

\begin{figure}[tbp]
\centering
\includegraphics[scale=0.45]{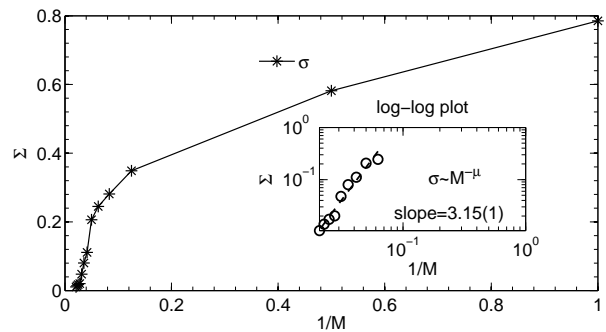}
\caption{Relative deviation $\Sigma$ from the multi-${\rm D}_2$ {\it Ansatz} is displayed as a function of $1/M$ with parameters $J=0.1, \phi=0.28$ and $F=0.1$. The inset reveals the relationship $\Sigma \sim M^{\mu}$ on a log-log scale, where the dashed line represents a power-law fit.}
\label{phi_diff_M}
\end{figure}
We extend our discussion to off-diagonal coupling, which is a formidable problem due to the intrinsic difficulties in achieving reliable results.
The off-diagonal coupling was emphasized as modulations of electron-electron interactions by ion vibrations in Mahan's
celebrated textbook on many-particle physics \cite{mahan}. Due to a lack of dependable solutions, a complete understanding of out-of-equilibrium dynamics for off-diagonal coupling remains elusive. Below we start with a validity check of our variational method.

Dynamics of the Holstein polaron with off-diagonal coupling under a constant external field is examined by using the multi-$\rm D_2$ {\it Ansatz}
with different multiplicity $M$. As shown in Figs.~\ref{exN8m1m8m20heom}(a)-(c) and (e), the time evolution of the exciton probability $P_{\rm ex}(t, n)$
is obtained by the $\rm D_2^{M=1}$, $\rm D_2^{M=8}$, $\rm D_2^{M=20}$ {\it Ans\"atze} and the HEOM method, respectively. Figs.~\ref{exN8m1m8m20heom}(b),(c)
and (e) display quite similar patterns and all three are largely different from Fig.~\ref{exN8m1m8m20heom}(a). Difference between $P_{\rm ex}(t, n)$
obtained by the $\rm D_2^{M=8}$ and $\rm D_2^{M=20}$ {\it Ansatz} as shown in Fig.~\ref{exN8m1m8m20heom}(d) is two orders of magnitude smaller than the
value of $P_{\rm ex}(t, n)$, pointing to the nearly converged results already obtained by $M=8$. Moreover, the difference in $P_{\rm ex}(t, n)$ between
the $\rm D_2^{M=20}$ {\it Ansatz} and the HEOM method, $\Delta P_{\rm ex}(t, n)$, as depicted in Fig.~\ref{exm1m16heom}(f), is also two orders of
magnitude smaller than the value of $P_{\rm ex}(t, n)$, showing the superior accuracy of the multi-D$_2$ {\it Ansatz}.

In addition, a quantity named the relative deviation $\Sigma$ (see Supporting Information) is also used to test the validity of our time-dependent variational approach by quantifying how closely the trial state follows the Schr\"odinger equation, as depicted in Fig.~\ref{phi_diff_M}. As the multiplicity $M$ increases, the relative deviation $\Sigma$ decreases and approaches zero as $M$ goes to infinity. This is supported by the relationship $\Sigma \sim M^{-\mu}$ with an exponent of $\mu=3.15(1)$ in the inset of Fig.~\ref{phi_diff_M}. Therefore, in the limit of large $M$, our variational method using the multi-$\rm D_2$ {\it Ansatz} provides a numerically exact solution to the Schr\"odinger equation in the presence of the external field.

\subsection{Anti-adiabatic regime}

\label{Weak and intermediate diagonal coupling}

The effect of a bosonic environment on BOs, i.e., BOs in a polaron framework, is further investigated in this subsection.  As usually envisioned for electronic
transport in crystals, motion of electrons is occasionally scattered by lattice vibrations \cite{th_70}. Pronounced modulations of the four-wave mixing signal with
characteristics of the temporal periodicity of BOs has been detected experimentally in an optical investigation of BOs in a semiconductor superlattice,
and can be attributed to lattice scattering \cite{fe_92}. Bouchard \textit{et al.} excluded interband transitions from being responsible for the signal
modulations and confirmed the single-band model as an appropriate approximation \cite{bo_95}. Several pioneering theoretical studies also demonstrated the
modulations of BOs by the electron-phonon interaction \cite{th_70, vi_11, lakhno_2004}. Despite tremendous experimental and theoretical efforts dedicated to the phonon
modulated BOs, the underlying mechanisms are still not well understood.

To be specific, both the acoustic and optical phonons have been experimentally revealed to coexist in the semiconductor superlattice \cite{kelvin_1986}, and amplitudes of exciton wave packets has been confirmed to be very sensitive to the precise excitation conditions in a weak external field \cite{ly_1997}. However, there is a lack of investigation on BOs dynamics influenced by the two phonon branches when the exciton starts from a Gaussian wave packet with varying widths.

\begin{figure}[tbp]
\centering
\includegraphics[scale=0.4]{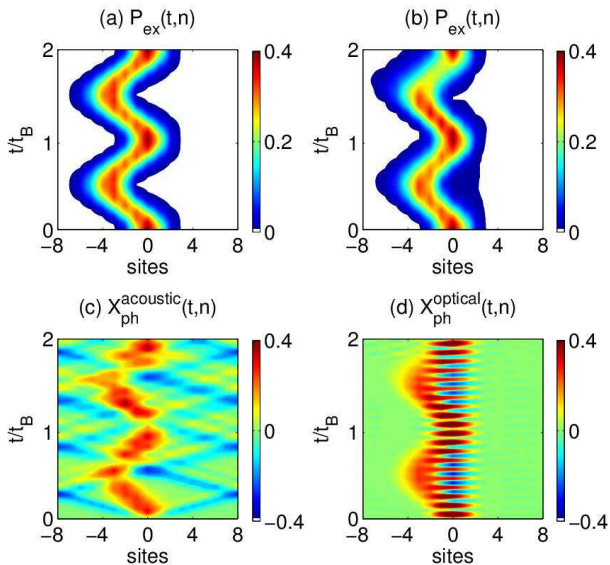}
\caption{Time evolution of the exciton probability $P_{\rm ex}(t, n)$ obtained from the $\rm D_2^{M=16}$ {\it Ansatz} in the case of $J=0.1$ and $F=0.1$ and an
initial broad Gaussian wave packet of $\sigma_0=1$ is displayed in (a) ($g=0$) and (b) ($g=0.4$). In the presence of weak diagonal coupling ($g=0.4$), the phonon displacement $X_{\rm ph}(t, n)$ is shown for (c) acoustic phonons and (d) optical phonons.}
\label{ex_d1_g0_g1_dif}
\end{figure}

\begin{figure}[tbp]
\centering
\includegraphics[scale=0.4]{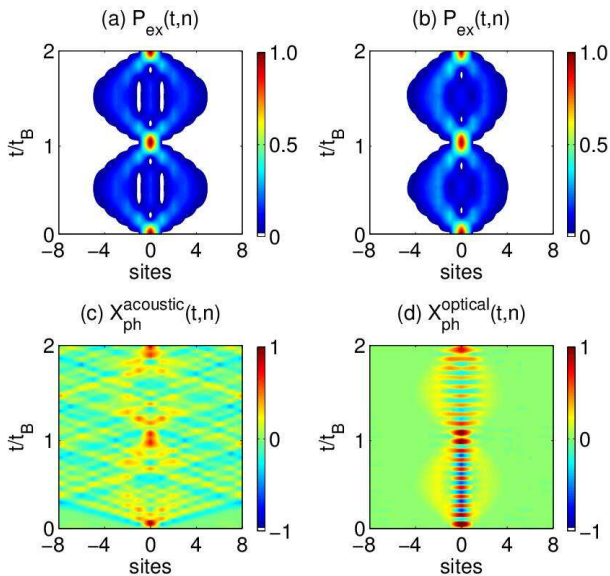}
\caption{Time evolution of the exciton probability $P_{\rm ex}(t, n)$ obtained from the $\rm D_2^{M=16}$ {\it Ansatz} in the case of $J=0.1$ and $F=0.1$ and an
initial narrow Gaussian wave packet of $\sigma_0=0.2$ is shown in (a) ($g=0$) and (b) ($g=0.4$). The phonon displacement $X_{\rm ph}(t, n)$ is shown for (c) the acoustic phonons and (d) optical phonons respectively in the presence of $g=0.4$.
}
\label{ex_d02_g0_g1_dif}
\end{figure}

In this subsection we will focus on dynamic properties of the Holstein polaron including the time evolution of the exciton probability and the phonon displacement for weak exciton-phonon coupling subject to a constant external electric field. Two typical scenarios are discussed by considering initial Gaussian wave packets
with initial widths $\sigma_0=1$ and $0.2$, as shown in Fig.~\ref{ex_d1_g0_g1_dif} and
Fig.~\ref{ex_d02_g0_g1_dif}, respectively. The initial Gaussian distribution \cite{st_2015} is described as follows
\begin{eqnarray}
\rho\left(n,t=0\right)	=	\left[\left(2\pi\right)^{1/2}\sigma_0\right]^{-1}\exp\left(-n^{2}/2\sigma_0^{2}\right)
\end{eqnarray}
which is centered at site $n=0$, and
the associated initial wave function is taken as the square root of the distribution:
\begin{eqnarray}
\psi\left(n,t=0\right)=\left[\left(2\pi\right)^{1/2}\sigma_0\right]^{-1/2}\exp\left(-n^{2}/4\sigma_0^{2}\right)
\end{eqnarray}
In particular, the roles played by the acoustic and optical phonons on the exciton transport are explicitly compared under the same external field. It is difficult to partition these branches rigorously \cite{bredas_2012}, and in order to avoid the complexity induced by partitioning these branches, we include only one phonon branch in each separate calculation and then compare the results of two calculations.

We start from using an initial broad Gaussian wave packet of $\sigma_0=1$. Transfer integral $J=0.1$ and an external field of $F=0.1$ is set in the calculations of this subsection.  In the absence of exciton-phonon coupling, as shown in Fig.~\ref{ex_d1_g0_g1_dif}(a), the exciton exhibits typical BOs, with the center of mass of the wave packet oscillates while its shape is essentially unchanged, in agreement with earlier theoretical work based on an ideal GaAs/Al${_x}$Ga$_{1-x}$As superlattice in a uniform electric field as described by a conventional flat-band picture \cite{st_2015, bo_95}. In the presence of the weak diagonal coupling ($g=0.4$), we focus on the anti-adiabatic regime where the phonon frequencies are larger than the transfer integral J. Only $P_{\rm ex}(t, n)$ of optical phonons is shown in Fig.~\ref{ex_d1_g0_g1_dif}(b) because the effect of the two phonon branches turned out to be similar. By comparing Figs.~\ref{ex_d1_g0_g1_dif}(a) and (b), the time periods of the exciton transport are the same because the temporal periodicity is determined by the external field. This agrees with the experimental observation that the detected signal is found to be modulated over time but the time period of the signal is found to be equal to the temporal periodicity of BOs \cite{fe_92}. The largest oscillation amplitude of the center of mass of the exciton wave packet in Fig.~\ref{ex_d1_g0_g1_dif}(a) is in accordance with the theoretical value of $4J/F$ \cite{ha_04}.

\begin{figure}[tbp]
\centering
\includegraphics[scale=0.46]{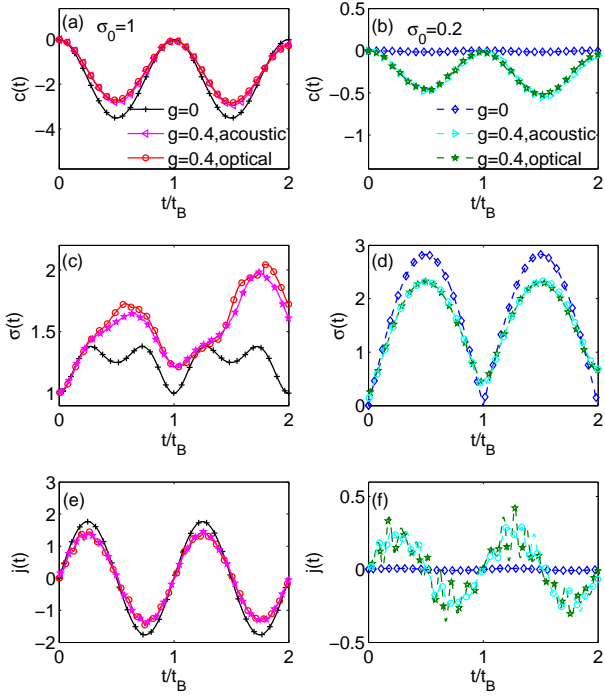}
\caption{ (a) Mean value $c(t)$, (c) standard deviation $\sigma(t)$ and (e) current $j(t)$ as functions of the time $t$ in the case of $J=0.1$  are
displayed using the initial standard deviation $\sigma_0=1$ and (b),(d),(f) using $\sigma_0=0.2$, respectively. In each panel, the results with $g=0$ and $g=0.4$ are compared.
}
\label{six}
\end{figure}

Even though the temporal periodicity is preserved after the exciton-phonon coupling is turned on, the spatial periodicity is changed over time. The
addition of the exciton-phonon coupling moves the center of mass of the exciton wave packet of $g=0$ closer to the initial location, which is in line
with the contrast in the mean value $c(t)$ between $g=0$ and $0.4$ in Fig.~\ref{six}(a). Not only the
motion of the center of mass of the exciton wave packet is changed, but also the width of the wave packet is enlarged as another characteristics of an effect of the
weak coupling, giving rise to a broadened wave packet. The width $\sigma(t)$ in Fig.~\ref{six}(c) is increased due to  smearing of the wave packet.

The phonon displacement $X_{\rm ph}(t, n)$ is shown in Figs.~\ref{ex_d1_g0_g1_dif}(c) and (d) for the acoustic and optical phonons, respectively. Before the exciton creation at $t=0$, the phonons are in their vacuum states for both scenarios. By comparing the two figures, parts of $X_{\rm ph}(t, n)$ (red and yellow) are found to propagate with BOs frequency $\omega_B$ because they are generated by the moving exciton wave packet and would in turn smear out the exciton wave fronts. As for the remainder of $X_{\rm ph}(t, n)$ (blue), a V-shaped feature and an oscillatory component with the phonon frequency $\omega_0$ are respectively found in Figs.~\ref{ex_d1_g0_g1_dif}(c) and (d). Moreover, the existence of ten peaks in one Bloch period $t_B$ can be attributed to the ratio $\omega_0/\omega_B$ (phonon frequency $\omega_0$ over BOs frequency). In the presented cases, $\omega_0$ is ten times of $\omega_B=de|F|/\hbar$, where $d$
is the lattice constant, $e$ is the charge of the exciton, and $d=e=\hbar=1$ is set. Meanwhile, the weak coupling affects the transport and the exciton
current in Fig.~\ref{six}(e). At zero coupling, the exciton current $j(t)$ (see Supporting Information) in the case of an initial broad Gaussian wave packet is $j(t)=2\sin Ft$ with the largest amplitude among all cases studied. The amplitude of the exciton current is decreased after the coupling is switched on, mitigating unidirectional exciton transport.

\begin{figure}[tbp]
\centering
\includegraphics[scale=0.42]{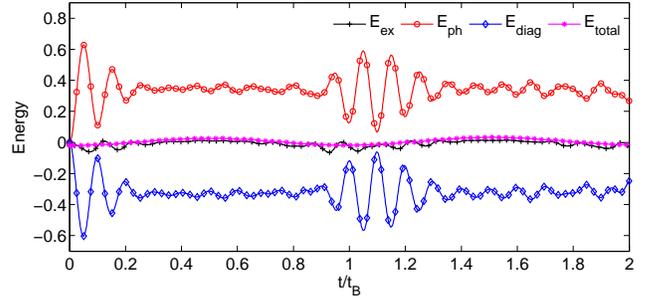}
\caption{$E_{\rm ex}(t), E_{\rm ph}(t), E_{\rm diag}(t)$ and $E_{\rm diag}(t)$ obtained by the $\rm D_2^{M=16}$ {\it Ansatz} in the case of $J=0.1, F=0.1$ and $\sigma_0=0.1$ in the presence of weak
diagonal coupling ($g=0.4$) with the optical phonons.}
\label{energyd02g01}
\end{figure}

Next, we consider the effect the weak exciton-phonon coupling on the dynamics with an initial narrow Gaussian wave packet of $\sigma_0=0.2$. As shown in Fig.~\ref{ex_d02_g0_g1_dif}(a), the exciton undergoes a symmetric breathing mode at zero coupling. The exciton wave packet propagates with its center of mass fixed at the original location and its width oscillates with the Bloch period, in accord with previous studies on the breathing mode \cite{ly_1997, bo_95, do_10}. After the addition of the weak diagonal coupling $g=0.4$, the exciton amplitude becomes larger in the left branch than that in the right branch (Fig.~\ref{ex_d02_g0_g1_dif}(b)) and the center of mass of the exciton
wave packet is moved away from the original location. These features of the exciton indicate that the symmetry of the breathing mode is broken by even minute exciton-phonon
interactions, and are further reflected by the mean value $c(t)$ plotted in Fig.~\ref{six}(b), which oscillates for the weak coupling
case of $g=0.4$ in contrast to being zero at all times for $g=0$. Recently, the breakdown of the breathing mode has been confirmed in semiconductor superlattices
\cite{di_94} and cold atoms trapped in optical lattice \cite{Trombettoni2001, ga_09, ga_11}, making our results here highly relevant in the study of breathing modes in those systems.
Furthermore, the width of the exciton wave packet for the weak coupling case of $g=0.4$ is reduced in comparison to that for $g=0$, as depicted in Fig.~\ref{six}(d).
A comparison between Figs.~\ref{six}(c) and (d) reveals that weak diagonal coupling has opposite influences on the width of the exciton wave packet: the exciton wave packet is broadened in the case of $\sigma_0=1$ while that in the case of $\sigma_0=0.2$ is suppressed after the exciton-phonon coupling is switched on.

Similar to the scenario of $\sigma_0=1$, the phonon displacement $X_{\rm ph}(t, n)$ of $\sigma_0=0.2$ propagates along with the movement of the exciton. As shown in Fig.~\ref{ex_d02_g0_g1_dif}(c), $X_{\rm ph}(t, n)$ maintains its V-shaped feature in the case of acoustic phonons. As for Fig.~\ref{ex_d02_g0_g1_dif}(d), characteristic oscillations with frequency $\omega_0$ is found in the case of optical phonons. Reciprocally, the generated $X_{\rm ph}(t, n)$ induces smearing of the exciton in its center. A brief comparison between Fig.~\ref{ex_d1_g0_g1_dif} and Fig.~\ref{ex_d02_g0_g1_dif} reveals that the exciton triggered phonon displacement leads to the weakening of the exciton wave packet in its center and edge, respectively. As shown in Fig.~\ref{six}(f), the current in the zero coupling case is zero due to the spatial symmetry of the exciton wave packet. The exciton current in the weak coupling exhibits fast beating with the characteristic frequency $\omega_0$ superimposed by slower BOs.

So far, we have focused on the impact of weak exciton-phonon coupling on the quantum transport with respect to the two typical scenarios. In general, BOs
are very sensitive to any kind of dephasing generated by the electron-hole Coulomb interaction effects or lattice imperfections, since they rely on the
coherent reflection of waves \cite{ga_11}. The electron-hole Coulomb interaction is pointed out to destroy the breathing mode by Dignam \textit{et al.}
after they examined the nature of the exciton wave packets in undoped semiconductor superlattices in a uniform along-axis electric field \cite{di_94}. The
effects of atom-atom interactions in the dilute BEC trapped in a periodic potential were actively studied using a discrete nonlinear Schr\"odinger
equation (DNLSE), and the atom-atom interactions can result in the decoherence of atomic BOs \cite{Trombettoni2001, kolovsky_09, ko_10} and enhancement or
suppression of the breathing width \cite{ga_09, ga_11}. To the best of our knowledge, ours is the
first investigation on the influence of the exciton-phonon interaction on both the typical BOs and the breathing mode.

Finally, we analyze the characteristics of the nonequilibrium dynamics of the ring system, including a comparison with the steady state of an infinite
linear chain. First of all, in the absence of an external electric field, in the anti-adiabatic limit of $\omega_0 \gg J$, exciton dynamics in both weak
and strong coupling regimes is dominated by coherent oscillations and negligible energy transfer, leaving a conserved total energy \cite{do_15}. After
the an external field is applied, the total energy is not conserved as plotted in Fig.~\ref{energyd02g01} because of the addition of the extra energy
acquired from the external field. Secondly, a steady state with a saturated current can be obtained in a linear chain that extends to infinity in both
directions, where the exciton acquires kinetic energy from the applied field while dissipating its energy by a net emission of optical phonons
\cite{th_70}. Being a steady state situation, the expectation value of the rate of change of electron momentum is zero, and the steady state is left with
an equation which balances the applied field, or rate of increase of electron momentum, against the rate of loss of momentum due to the lattice
scattering. This result is an explicit field dependence in terms of the steady-state velocity of the electron. The key feature of the steady state
is also explored in Ref.~[\citenum{vi_11, lakhno_2010, lakhno_2011}]. Equivalently, the condition of the steady state can be that there exist linear regimes of the total energy and phonon energy following the equality of $dE_{\rm total}/dt=dE_{\rm ph}/dt$ \cite{vi_11}, meaning that the total energy gain is entirely absorbed by the lattice. Specifically, with increasing time $t$, the total energy vs time approaches a straight line, and the phonon energy has a linear time dependence as well. But in our ring system, the condition of the steady state can not be met as this equation can not be satisfied from the curves of $E_{\rm total}$ and $E_{\rm ph}$ as shown in
Fig.~\ref{energyd02g01}. For example, the value of $dE_{\rm ph}/dt$ at $t=2t_B$ is not equal to but $9$ times of the value of $dE_{\rm total}/dt$. The
energy imported by the external field can not cancel out the energy emitted to the phonons, leaving the exciton energy oscillates over time. Thus, the
electron's acceleration in the ring can no longer be balanced out by the lattice's deceleration and the electron can not drift with a saturated velocity. Consequently, no saturated constant current is found in the anti-adiabatic regime for the small ring systems, and steady states in the adiabatic regime will be investigated elsewhere.

To sum up, weak exciton-phonon coupling in the anti-adiabatic regime breaks the spatial periodicity but retains the temporal periodicity of both the typical BOs and the breathing mode. The exciton movement shows similar features in spite of the difference between the acoustic and optical phonon branches. In fact, in a weak external field, when diagonal coupling is strong between the exciton and both the acoustic and optical phonons, the exciton become localized, leading to the diminishing of the typical BOs and the breathing mode. The localization is similar to that in the moderate external field cases in the next subsection, therefore the details are not shown. We also show that no saturated constant current is found in the ring system for the duration of our simulation.

\subsection{Strong diagonal coupling}

\label{Strong diagonal coupling}

\begin{figure}[tbp]
\centering
\includegraphics[scale=0.42]{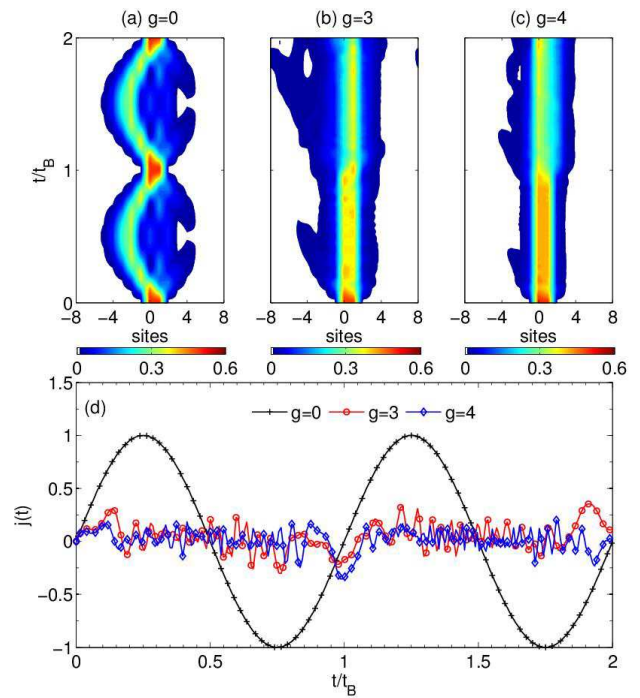}
\caption{Time evolution of the exciton probability $P_{\rm ex}(t, n)$ obtained from the $\rm D_2^{M=16}$ {\it Ansatz} in the case of $J=1, N=16$ and a
moderate external field of $F=1$ is displayed for different diagonal coupling strengths: (a) $g=0$, (b) $g=3$ and (c) $g=4$. Corresponding currents $j(t)$ are shown in (d).}
\label{F1}
\end{figure}

\begin{figure}[tbp]
\centering
\includegraphics[scale=0.42]{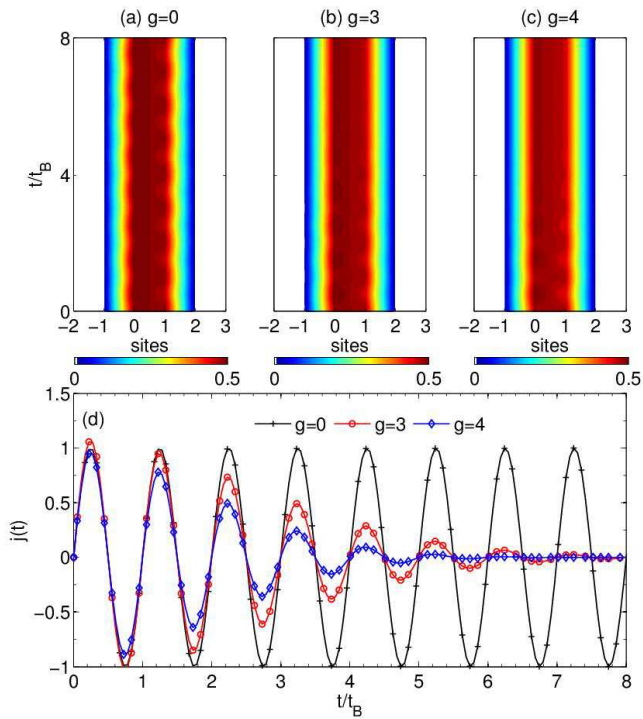}
\caption{Time evolution of the exciton probability $P_{\rm ex}(t, n)$ obtained from the $\rm D_2^{M=16}$ {\it Ansatz} in the case of $J=1, N=16$ and a
strong external field of $F=70$ is displayed for different diagonal coupling strengths: (a) $g=0$, (b) $g=3$ and (c) $g=4$. Corresponding currents $j(t)$ are shown in (d).}
\label{F70}
\end{figure}

In this subsection we seek to present a general picture of the exciton wave packet evolution in the regime of the strong diagonal coupling. We first
investigate the exciton wave packets as well as the exciton currents under a moderate external field of $F=1$. The time evolution of the exciton probability
$P_{\rm ex}(t, n)$ for different diagonal coupling strengths $g=0$, $3$ and $4$ is
illustrated in Figs.~\ref{F1}(a), (b) and (c), respectively. The
exciton is created on two nearest neighbouring sites. We set $J=1$ and $ N=16$. As shown in Figs.~\ref{F1}(a)-(c), the bare exciton exhibits a partially BOs pattern, while the strong exciton-phonon coupling is found to localize the
exciton to the initial excitation sites. BOs are largely quenched due to strong exciton-phonon coupling \cite{lakhno_2007}. This behaviour is also demonstrated in Fig.~\ref{F1}(d), which shows that the amplitude of the exciton current decreases with increasing
exciton-phonon coupling and fluctuates around zero.

We then discuss the case of a strong external field. Using the same parameters as the case of $F=1$, the time evolution of the exciton probability $P_{\rm ex}(t, n)$ and exciton current $j(t)$ under a strong field of $F=70$ are presented in Figs.~\ref{F70}(a)-(c) and Fig.~\ref{F70}(d), respectively. Due to the strong external field, the exciton is found to be localized at the initial site of excitation irrespective of the exciton-phonon coupling strength. A bare exciton case is shown in Fig.~\ref{F70}(a) for comparison. The localization of the exciton wave packet can be rationalized in terms of Wannier-Stark states \cite{wa_59}. In the Wannier representation, the Wannier-Stark states are denoted as $|\Psi_n\rangle=\sum_mJ_{m-n}(\gamma)|m\rangle$, where $J_{m-n}(\gamma)$ is the Bessel
function of order $m-n$ with $\gamma=2J/Fd$ \cite{ha_04}. Due to properties of the Bessel functions, the Wannnier-Stark states extend over
the interval $L\simeq2J/F$, which leads to well localized states in the limit of a strong external field \cite{Emin1987}. As shown in Figs.~\ref{F70}(b) and (c), strong exciton-phonon coupling leads to the decay of BOs. The exciton currents in the presence of strong exciton-phonon coupling
exhibits damped oscillations with a time scale of $8t_B$ ($6t_B$) for $g=3$ ($4$).

\section{Conclusion}
\label{Conclusions}

We have studied transient dynamics of the Holstein polaron in a one-dimensional ring under a constant external field using
the Dirac-Frenkel time-dependent variational principle and the novel multi-D$_2$ {\it Ansatz}, which is a linear combination of the usual Davydov $\rm D_2$ trial states. In both the diagonal and off-diagonal coupling cases, our efficient variational calculations are in perfect agreement with those
obtained from the numerically exact HEOM method. Moreover, the relative deviation is found to decay with the increasing multiplicity of the multi-D$_2$ {\it Ansatz}, which vanishes in the limit of $M \to \infty$, inferring that our approach is numerically exact in that limit.

Firstly, the influence of the initial condition is studied in the absence of exciton-phonon coupling. For an initial broad Gaussian wave packet,
typical BOs are found with the center of the wave packet oscillating but its shape essentially unchanged. Starting from an initial narrow Gaussian wave packet, the exciton wave exhibits a symmetric breathing mode with its width oscillating with the Bloch period and its center of mass fixed at the original location.

The effect of the exciton-phonon coupling is the focus of our investigation. In general, weak diagonal coupling breaks the spatial periodicity while
keeping the temporal periodicity of the exciton wave. For an initial broad Gaussian wave packet, the application of weak diagonal coupling modifies BOs with the exciton wave packet broadened and the exciton current reduced. For an initial narrow Gaussian wave packet, after the addition of weak coupling,
the spatial symmetry of the exciton wave is broken and the center of mass of the exciton wave packet oscillates away from the original position, leading to a
non-zero exciton current. In particular, a saturated current which exists in an infinite linear lattice is not found in a finite-sized ring within the anti-adiabatic regime, since the energy
imported by the external field is not entirely absorbed by the lattice, leaving the steady state unreachable.

For strong exciton-phonon coupling, the variational method using the multi-D$_2$ {\it Ansatz} is found to be highly accurate while it is prohibitively
expensive for the HEOM method to tackle higher phonon excited states. The exciton wave packet is found to be localized due to either strong diagonal coupling
or a strong external field. Finally, strong diagonal coupling gives rise to the decay of BOs under a strong external field, leading to damped
oscillations of the exciton current.

\section*{Acknowledgments}
Support from the Singapore National Research Foundation through the Competitive Research Programme (CRP) under Project No.~NRF-CRP5-2009-04 is gratefully
acknowledged.

\appendix
\section{A gauge transformed Hamiltonian}
\label{Transformation}
The Hamiltonian without the phonon part in infinite lattice can be written in two parts:
the tight-binding Hamiltonian
\begin{eqnarray}
&&H_{0}=-J\sum_{n}\left(a_{n+1}^{\dagger}a_{n}+a_{n}^{\dagger}a_{n+1}\right),
\end{eqnarray}
scalar potential interaction
\begin{eqnarray}
&&H_{int}=F\sum_{n}a_{n}^{\dagger}na_{n},
\end{eqnarray}
we can introduce the vector potential $A\left(t\right)=-Ft$
and employ the gauge transformation
\begin{eqnarray}
&&\tilde{\psi}\left(t\right)	=	e^{-iA\left(t\right)x}\psi\left(t\right)
\end{eqnarray}
where $x=\sum_{n}a_{n}^{\dagger}na_{n}$ is the position operator.
\begin{eqnarray}
&&i\frac{\partial\psi\left(t\right)}{\partial t}=\left(H_{0}+H_{int}\right)\psi\left(t\right)\nonumber\\
&&i\frac{\partial\tilde{\psi}\left(t\right)}{\partial t}=\tilde{H}\tilde{\psi}\left(t\right)
\end{eqnarray}
can be satisfied, and spatial translational symmetry can be restored in $\tilde{H}$.

The form of $\tilde{H}$ is shown in the following.		
\begin{eqnarray}
&&i\frac{\partial\tilde{\psi}\left(t\right)}{\partial t}\nonumber\\
&&	=	i\frac{\partial\left(-iA\left(t\right)x\right)}{\partial
t}e^{-iA\left(t\right)x}\psi\left(t\right)+e^{-iA\left(t\right)x}i\frac{\partial\psi\left(t\right)}{\partial t}\nonumber\\
&&	=	-Fxe^{-iA\left(t\right)x}\psi\left(t\right)+e^{-iA\left(t\right)x}i\frac{\partial\psi\left(t\right)}{\partial t}\nonumber\\
&&	=	-H_{int}e^{-iA\left(t\right)x}\psi\left(t\right)+e^{-iA\left(t\right)x}\left(H_{0}+H_{int}\right)\psi\left(t\right)\nonumber\\
&&	=	e^{-iA\left(t\right)x}H_{0}\psi\left(t\right)
\end{eqnarray}
\begin{eqnarray}
&&\tilde{H}\tilde{\psi}\left(t\right)	=	\tilde{H}e^{-iA\left(t\right)x}\psi\left(t\right)
\end{eqnarray}
\begin{eqnarray}
&&e^{-iA\left(t\right)x}H_{0}\psi\left(t\right)	=	\tilde{H}e^{-iA\left(t\right)x}\psi\left(t\right)\nonumber\\
&&e^{-iA\left(t\right)x}H_{0}	=	\tilde{H}e^{-iA\left(t\right)x}
\end{eqnarray}
 both side left multiply by $e^{iA\left(t\right)x}$
\begin{eqnarray}
&&\tilde{H}	=	e^{-iA\left(t\right)x}H_{0}e^{iA\left(t\right)x}
\end{eqnarray}
 using the following rule
\begin{eqnarray}
&&e^{\hat{A}}\hat{B}e^{-\hat{A}}	=	 \hat{B}+\left[\hat{A},\hat{B}\right]+\frac{1}{2!}\left[\hat{A},\left[\hat{A},\hat{B}\right]\right]\nonumber\\
&&		+\frac{1}{3!}\left[\hat{A},\left[\hat{A},\left[\hat{A},\hat{B}\right]\right]\right]+...
\end{eqnarray}
 make
\begin{eqnarray}
&&\hat{A}	=	-iA\left(t\right)x=-iA\left(t\right)\sum_{n}a_{n}^{\dagger}na_{n}\nonumber\\
&&\hat{B}	=	\frac{H_{0}}{-J}=\sum_{n}\left(a_{n+1}^{\dagger}a_{n}+a_{n}^{\dagger}a_{n+1}\right)
\end{eqnarray}
we can obtain
\begin{eqnarray}
&& 		\left[\hat{A},\hat{B}\right]\nonumber\\
&&	=	-iA\left(t\right)\sum_{n'}a_{n'}^{\dagger}n'a_{n'}\left[\sum_{n}\left(a_{n+1}^{\dagger}a_{n}+a_{n}^{\dagger}a_{n+1}\right)\right]\nonumber\\
&&		
-\left[\sum_{n}\left(a_{n+1}^{\dagger}a_{n}+a_{n}^{\dagger}a_{n+1}\right)\right]\left[-iA\left(t\right)\sum_{n'}a_{n'}^{\dagger}n'a_{n'}\right]\nonumber\\
&&	=	-iA\left(t\right)\left[\sum_{n}\left(n+1\right)a_{n+1}^{\dagger}a_{n}+\sum_{n}na_{n}^{\dagger}a_{n+1}\right]\nonumber\\
&&		+iA\left(t\right)\left[\sum_{n}a_{n+1}^{\dagger}na_{n}+\sum_{n}a_{n}^{\dagger}\left(n+1\right)a_{n+1}\right]\nonumber\\
&&	=	\left(-iA\left(t\right)\right)\sum_{n}a_{n+1}^{\dagger}a_{n}+\left(iA\left(t\right)\right)\sum_{n}a_{n}^{\dagger}a_{n+1}
\end{eqnarray}

\begin{eqnarray}
&&\left[\hat{A},\left[\hat{A},\hat{B}\right]\right]\nonumber\\
&&=\left(-iA\left(t\right)\right)^{2}\sum_{n}a_{n+1}^{\dagger}a_{n}+\left(iA\left(t\right)\right)^{2}\sum_{n}a_{n}^{\dagger}a_{n+1}
\end{eqnarray}

\begin{eqnarray}
&&\left[\hat{A},\left[\hat{A},\left[\hat{A},\hat{B}\right]\right]\right]\nonumber\\
&&=\left(-iA\left(t\right)\right)^{3}\sum_{n}a_{n+1}^{\dagger}a_{n}+\left(iA\left(t\right)\right)^{3}\sum_{n}a_{n}^{\dagger}a_{n+1}
\end{eqnarray}
thus
\begin{eqnarray}
&&e^{-iA\left(t\right)x}\frac{H_{0}}{-J}e^{iA\left(t\right)x}\nonumber\\
&&=	\left(\sum_{n}\left(a_{n+1}^{\dagger}a_{n}+a_{n}^{\dagger}a_{n+1}\right)\right)\nonumber\\
&&	 +\left(\left(-iA\left(t\right)\right)\sum_{n}a_{n+1}^{\dagger}a_{n}+\left(iA\left(t\right)\right)\sum_{n}a_{n}^{\dagger}a_{n+1}\right)\nonumber\\
&&	
+\frac{1}{2!}\left(\left(-iA\left(t\right)\right)^{2}\sum_{n}a_{n+1}^{\dagger}a_{n}+\left(iA\left(t\right)\right)^{2}\sum_{n}a_{n}^{\dagger}a_{n+1}\right)\nonumber\\
&&	
+\frac{1}{3!}\left(\left(-iA\left(t\right)\right)^{3}\sum_{n}a_{n+1}^{\dagger}a_{n}+\left(iA\left(t\right)\right)^{3}\sum_{n}a_{n}^{\dagger}a_{n+1}\right)\nonumber\\
&&	+...\nonumber\\
&&=	\sum_{n}a_{n+1}^{\dagger}a_{n}e^{-iA\left(t\right)x}+\sum_{n}a_{n}^{\dagger}a_{n+1}e^{iA\left(t\right)x}
\end{eqnarray}
therefore
\begin{eqnarray}
&&\tilde{H}	=	-J\left[\sum_{n}a_{n+1}^{\dagger}a_{n}e^{-iA\left(t\right)x}+\sum_{n}a_{n}^{\dagger}a_{n+1}e^{iA\left(t\right)x}\right]\nonumber\\
&&	=	-J\left(\sum_{n}a_{n+1}^{\dagger}a_{n}e^{iFt}+\sum_{n}a_{n}^{\dagger}a_{n+1}e^{-iFt}\right)\nonumber\\
&&	=	-J\sum_{n}a_{n}^{\dagger}\left(e^{-iFt}a_{n+1}+e^{iFt}a_{n-1}\right)
\end{eqnarray}

\section{The Multi-${\rm D}_2$ trial state}
\label{Equations}
The energies of the system is shown in the following,
\begin{eqnarray}
&&\left\langle D_{2}^M\left(t\right)\right|H_{ex}\left|D_{2}^M\left(t\right)\right\rangle 	=	\nonumber \\
&&-J\sum_{i,j}^{M}\sum_{n}\psi_{jn}^{\ast}\left(e^{-iFt}\psi_{i,n+1}+e^{iFt}\psi_{i,n-1}\right)S_{ji}\nonumber \\
&&\left\langle D_{2}^M\left(t\right)\right|H_{ph}\left|D_{2}^M\left(t\right)\right\rangle   =	\nonumber \\
&&\sum_{i,j}^{M}\sum_{n}\psi_{jn}^{\ast}\psi_{in}\sum_{q}\omega_{q}\lambda_{jq}^{\ast}\lambda_{iq}S_{ji}\nonumber \\
&&\left\langle D_{2}^M\left(t\right)\right|H_{ex-ph}^{diag}\left|D_{2}^M\left(t\right)\right\rangle 	=	\nonumber \\
&&-N^{-1/2}g\sum_{i,j}^{M}\sum_{n}\psi_{jn}^{\ast}\psi_{in}\sum_{q}\left(e^{iqn}\lambda_{iq}+e^{-iqn}\lambda_{jq}^{\ast}\right)S_{ji}\nonumber \\
&&\left\langle {\rm D}^M_{2}\left(t\right)\right|H_{ex-ph}^{o.d.}\left|{\rm D}^M_{2}\left(t\right)\right\rangle=\frac{1}{2}N^{-1/2}\phi\sum_{n,q}\sum_{i,j}^{M}\omega_{q}S_{ji}\nonumber \\
&& \times\{\psi_{jn}^{\ast}\psi_{i,n+1}[e^{iqn}(e^{iq}-1)\lambda_{iq}+e^{-iqn}(e^{-iq}-1)\lambda_{jq}^{\ast}]\nonumber \\
&& +\psi_{jn}^{\ast}\psi_{i,n-1}[e^{iqn}(1-e^{-iq})\lambda_{iq}+e^{-iqn}(1-e^{iq})\lambda_{jq}^{\ast}]\}\nonumber \\
\end{eqnarray}

The Dirac-Frenkel variational principle results in equations of motion including
\begin{eqnarray}
&&		-i\sum_{i}\dot{\psi}_{in}S_{ki}\nonumber \\	
&&-\frac{i}{2}\sum_{i}\psi_{in}\sum_{q}\left(2\lambda_{kq}^{\ast}\dot{\lambda}_{iq}-\dot{\lambda}_{iq}\lambda_{iq}^{\ast}-\lambda_{iq}\dot{\lambda}_{iq}^{\ast}\right)S_{k,i}\nonumber
\\
&&	=	J\sum_{i}\left(e^{-iFt}\psi_{i,n+1}+e^{iFt}\psi_{i,n-1}\right)S_{ki}\nonumber \\
&&		-\sum_{i}\psi_{in}\sum_{q}\omega_{q}\lambda_{kq}^{\ast}\lambda_{iq}S_{ki}\nonumber \\
&&		+N^{-1/2}g\sum_{i}\psi_{in}\sum_{q}\left(e^{iqn}\lambda_{iq}+e^{-iqn}\lambda_{kq}^{\ast}\right)S_{ki}\nonumber \\
&& -\frac{1}{2}N^{-1/2}\phi\sum_{i}\sum_{q}\omega_{q}\psi_{i,n+1}[e^{iqn}(e^{iq}-1)\lambda_{iq}\nonumber \\
&&+e^{-iqn}(e^{-iq}-1)\lambda_{kq}^{\ast}]S_{ki}\nonumber \\
&& -\frac{1}{2}N^{-1/2}\phi\sum_{i}\sum_{q}\omega_{q}\psi_{i,n-1}[e^{iqn}(1-e^{-iq})\lambda_{iq}\nonumber \\
&&+e^{-iqn}(1-e^{iq})\lambda_{kq}^{\ast}]S_{ki}, \nonumber \\
\end{eqnarray}
and
\begin{eqnarray}
&&		 -i\sum_{i}\sum_{n}\psi_{kn}^{\ast}\dot{\psi}_{in}\lambda_{iq}S_{ki}-i\sum_{i}\sum_{n}\psi_{kn}^{\ast}\psi_{in}\dot{\lambda}_{iq}S_{ki}\nonumber
\\
&&-\frac{i}{2}\sum_{i}\sum_{n}\psi_{kn}^{\ast}\psi_{in}\lambda_{iq}S_{k,i}\sum_{p}\left(2\lambda_{kp}^{\ast}\dot{\lambda}_{ip}-\dot{\lambda}_{ip}\lambda_{ip}^{\ast}-\lambda_{ip}\dot{\lambda}_{ip}^{\ast}\right)\nonumber \\
&&		=J\sum_{i}\sum_{n}\psi_{kn}^{\ast}\left(e^{-iFt}\psi_{i,n+1}+e^{iFt}\psi_{i,n-1}\right)\lambda_{iq}S_{k,i}\nonumber \\
&&		-\sum_{i}\sum_{n}\psi_{kn}^{\ast}\psi_{in}\left(\omega_{q}+\sum_{p}\omega_{p}\lambda_{kp}^{\ast}\lambda_{ip}\right)\lambda_{iq}S_{ki}\nonumber \\
&&+N^{-1/2}g\sum_{i}\sum_{n}\psi_{kn}^{\ast}\psi_{in}e^{-iqn}S_{ki}\nonumber \\
&&		+N^{-1/2}g\sum_{i}\sum_{n}\psi_{kn}^{\ast}\psi_{in}\lambda_{iq}\sum_{p}\left(e^{ipn}\lambda_{ip}+e^{-ipn}\lambda_{kp}^{\ast}\right)S_{k,i}\nonumber \\
&&-\frac{1}{2}N^{-1/2}\phi\sum_{n}\sum_{i}\omega_{q}\psi_{kn}^{\ast}[\psi_{i,n+1}e^{-iqn}(e^{-iq}-1)\nonumber\\
&&+\psi_{i,n-1}e^{-iqn}(1-e^{iq})]S_{ki}\nonumber \\
&&-\frac{1}{2}N^{-1/2}\phi\sum_{n}\sum_{i}\left(\psi_{k,n+1}^{\ast}\psi_{i,n}+\psi_{kn}^{\ast}\psi_{i,n+1}\right)\lambda_{iq}\nonumber\\
&&\sum_{p}\omega_{p}[e^{ipn}(e^{ip}-1)\lambda_{ip}+e^{-ipn}(e^{-ip}-1)\lambda_{kp}^{\ast}]S_{k,i},\nonumber\\
\end{eqnarray}
the current of the system under external field is described as,
\begin{eqnarray}
&&j\left(t\right)	=	\left\langle \hat{I}\left(t\right)\right\rangle,\nonumber \\
\end{eqnarray}
\begin{eqnarray}
&&\hat{I}\left(t\right)	=	i\left(\sum_{n}e^{-iFt}a_{n}^{\dagger}a_{n+1}-H.c.\right)\nonumber \\
&&	=	i\sum_{n}a_{n}^{\dagger}\left(e^{-iFt}a_{n+1}-e^{iFt}a_{n-1}\right),\nonumber \\
\end{eqnarray}
\begin{eqnarray}
&&j\left(t\right)	=	\left\langle
D_2^{M}\left(t\right)\right|i\sum_{n}\left(e^{-iFt}a_{n}^{\dagger}a_{n+1}-e^{iFt}a_{n}^{\dagger}a_{n-1}\right)\left|D_2^{M}\left(t\right)\right\rangle\nonumber \\
&&	=	
i\sum_{i,j}^{M}\sum_{n}\psi_{j,n}^{*}\left(t\right)\left[e^{-iFt}\psi_{i,n+1}\left(t\right)-e^{iFt}\psi_{i,n-1}\left(t\right)\right]S_{ji},\nonumber \\
\end{eqnarray}

\section{Relative deviation}
\label{relative deviation}
Assuming the trial wave function $|{\rm D}^M_{2}(t)\rangle=|\Psi(t)\rangle$ ($|\Psi(t)\rangle$ is the real wave function) at the time $t$, we introduce a deviation vector $\vec{\delta}(t)$ to quantify the accuracy of the variational dynamics based on the multiple Davydov trial states,
\begin{eqnarray}
\vec{\delta}(t) & =  & \vec{\chi}(t) -\vec{\gamma}(t)  \nonumber \\
& = & \frac{\partial}{\partial t}|\Psi(t)\rangle - \frac{\partial}{\partial t}|{\rm D}^M_{2}(t)\rangle.
\label{deviation_1}
\end{eqnarray}
where the vectors $\vec{\chi}(t)$ and $\vec{\gamma}(t)$ obey the Schr\"{o}dinger equation  $\vec{\chi}(t)=\partial |\Psi(t)\rangle / \partial t = \frac{1}{i}\hat{H}|\Psi(t)\rangle$ and the Dirac-Frenkel variational dynamics $\vec{\gamma}(t)=\partial |{\rm D}^M_{2}\rangle / \partial t$, respectively. Using the Schr\"{o}dinger equation and the relationship $|\Psi(t)\rangle=|{\rm D}^M_{1,2}(t)\rangle$ at the moment $t$, the deviation vector $\vec{\delta}(t)$ can be calculated as
\begin{equation}
\vec{\delta}(t) = \frac{1}{i}\hat{H}|{\rm D}^M_{2}(t)\rangle - \frac{\partial}{\partial t}|{\rm D}^M_{2}(t)\rangle.
\label{deviation_2}
\end{equation}
Thus, the deviation from the exact Schr{\"o}dinger dynamics can be indicated by the amplitude of the deviation vector $\Delta(t)=||\vec{\delta}(t)||$.
In order to view the deviation in the parameter space $(W,J,g,\phi)$, a dimensionless relative deviation $\Sigma$ is calculated as
\begin{equation}
\Sigma = \frac{{\rm max}\{\Delta(t)\} }{{\rm mean}\{N_{\rm err}(t)\}}, \quad \quad t \in [0, t_{\rm max}].
\label{relative_error}
\end{equation}
where $N_{\rm err}(t)=||\vec{\chi}(t)||$ is the amplitude of the time derivative of the wave function,
\begin{eqnarray}
N_{\rm err}(t) & = & \sqrt{-\langle\frac{\partial}{\partial t}\Psi(t)|\frac{\partial}{\partial t}\Psi(t)\rangle} \nonumber \\
& = & \sqrt{\langle {\rm D}^M_{2}(t)|\hat{H}^2|{\rm D}^M_{2}(t)\rangle}
\end{eqnarray}

\section{Hierarchy equation of motion}
\label{Hierarchy equation of motion}
The reduced density matrix element for the exciton system in the site basis can be written in the path
integral form with the factorized initial condition as
\begin{eqnarray}\label{Rho}
\rho{(n,n';t)}&=&\int\mathcal{D}n\int\mathcal{D}n'\rho{(n_0,n_0^{'};t_0)}\nonumber \\
&&\times{e^{iS[n;t]}}F(n,n';t)e^{-iS[n';t]}
\end{eqnarray}
where $S[n]$ is the action of the exciton system, and $F[n,n']$ is the Feynman-Vernon influence functional given by
\begin{eqnarray}
F[n,n']&=&\exp\{-\sum_q\omega_q^2\int_{t_0}^tds\int_{t_0}^sds'\nonumber \\
&&{V_q^{\ast}}^{\times}(s)\times\big[V_q^{\times}(s')\coth(\beta\omega_q/2)\cos(\omega_q(s-s'))\nonumber \\
&&-iV_q^{\circ}(s')\sin{(\omega_q(s-s'))}\big]\}
\end{eqnarray}
where $\beta$ is the inverse of temperature $(\beta=1/k_BT)$, and we define following abbreviations
\begin{eqnarray}
V_q^{\times}=V_q(n)-V_q(n') \\
V_q^{\circ}=V_q(n)+V_q(n')
\end{eqnarray}
with $V_q^{\ast}$ denoting the matrix representation of the operator $\hat{V}_q^{\dagger}=g\sum_n\hat{a}_n^{\dagger}\hat{a}_ne^{iqn}$

Taking derivative of Eq.~(\ref{Rho}),we have
\begin{eqnarray}
\frac{\partial}{\partial{t}}\rho(n,n';t)&&=-i\mathcal{L}\rho(n,n';t)\nonumber \\
&&-\sum_q\Phi_q(t)\int\mathcal{D}n\int\mathcal{D}n'\rho(n_0,n_0^{'};t_0)\int_0^tds'\nonumber \\
&&[e^{i\omega_q(t-s')}\Theta_{q+}(s')+e^{-i\omega_q(t-s')}\Theta_{q-}(s')]\nonumber \\
&&\times{e^{iS[n,t]}}F(n,n';t)e^{-iS[n',t]},
\end{eqnarray}
with the following super-operators defined as
\begin{eqnarray}
\hat{\Phi}_q(t)&=&\omega_q^2{V_q^{\dagger}}^{\times}(t)/2,\\
\hat{\Theta}_{q\pm}(t)&=&{V_q}^{\times}(t)\coth(\beta\omega_q/2)\mp{V_q}^{\circ}(t).
\end{eqnarray}

We then introduce the auxiliary operator $\rho_{m_{1\pm},m_{2\pm},\cdots,m_{N\pm}}(n,n';t)$ by its matrix element as
\begin{eqnarray}
\rho_{m_{1\pm},m_{2\pm},\cdots,m_{N\pm}}(n,n';t)=\int\mathcal{D}n\int\mathcal{D}n'\rho(n_0,n_0^{'};t_0)\prod_{q=1}^{N}\nonumber \\
(\int_{t_0}^tdse^{i\omega_q(t-s)}\Theta_{q+}(s))^{m_{q+}}(\int_{t_0}^{t}dse^{-i\omega_q(t-s)}\Theta_{q-}(s))^{m_{q-}}\nonumber \\
\times{e^{iS[n,t]}}F(n,n')e^{-iS[n',t]}\nonumber \\
\end{eqnarray}
for non-negative integers $m_{1\pm},m_{2\pm},...,m_{N\pm}$. Note that $\hat{\rho}_{0......0}(t)=\hat{\rho}(t)$ as well as other auxiliary density matrices
contain the complete information on the Liouville space wave packets. Differentiating $\rho_{m_{1\pm},m_{2\pm},...,m_{N\pm}}(n,n';t)$ with respect to $t$,
we can obtain a set of equations
\begin{eqnarray}\label{HEOM}
\frac{\partial}{\partial{t}}\hat{\rho}_{m_{1\pm},\cdots,m_{N\pm}}(t)=-i\mathcal{L}\hat{\rho}_{m_{1\pm},\cdots,m_{N\pm}}(t)\nonumber \\
-i\sum_q\omega_q(m_{q-}-m_{q+})\hat{\rho}_{m_{1\pm},\cdots,m_{N\pm}}(t)\nonumber \\
-\sum_q\hat{\Phi}_q(\hat{\rho}_{m_{1\pm},\cdots,m_{q+}+1,m_{q-},\cdots,m_{N\pm}}(t)+\nonumber \\
\hat{\rho}_{m_{1\pm},\cdots,m_{q+},m_{q-}+1,\cdots,m_{N\pm}}(t))\nonumber \\
+\sum_q(m_{q+}\hat{\Theta}_{q+}\hat{\rho}_{m_{1\pm},\cdots,m_{q+}-1,m_{q-},\cdots,m_{N\pm}}(t)+\nonumber \\
m_{q-}\hat{\Theta}_{q-}\hat{\rho}_{m_{1\pm},\cdots,m_{q+},m_{q-}-1,\cdots,m_{N\pm}}(t)).
\end{eqnarray}

It should be noted that the HEOM consists of an infinite number of equations, but they can be truncated at finite number of hierarchy elements by the
terminator for the practical calculation as
\begin{eqnarray}
\frac{\partial}{\partial{t}}\hat{\rho}_{m_{1\pm},\cdots,m_{N\pm}}(t)&=&-(i\mathcal{L}+i\sum_q\omega_q(m_{q-}-m_{q+}))\nonumber \\
&&\times{\hat{\rho}_{m_{1\pm},\cdots,m_{N\pm}}(t)},
\end{eqnarray}

---

\end{document}